# Nonlinear pulse propagation in InAs/InP quantum-dot optical amplifiers: Rabi-oscillations in the presence of non-resonant nonlinearities

O. Karni[*,1], A. K. Mishra[1], G. Eisenstein[1], and J. P. Reithmaier[2]

[1] Electrical Engineering Dept., Technion – Israel Institute of Technology, Haifa 32000, Israel.

[2] Technische Physik, Institute of Nanostructure Technologies and Analytics, CINSaT, University of Kassel, 34132 Kassel, Germany.

[*] e-mail: oulrik@tx.technion.ac.il

**Abstract**

We study the interplay between coherent light-matter interactions and non-resonant pulse propagation effects when ultra-short pulses propagate in room-temperature quantum-dot (QD) semiconductor optical amplifiers (SOAs). The signatures observed on a pulse envelope after propagating in a transparent SOA, when coherent Rabi-oscillations are absent, highlight the contribution of two-photon absorption (TPA), and its accompanying Kerr-like effect, as well as of linear dispersion, to the modification of the pulse complex electric field profile. These effects are incorporated into our previously developed finite-difference time-domain comprehensive model that describes the interaction between the pulses and the QD SOA. The present, generalized, model is used to investigate the combined effect of coherent and non-resonant phenomena in the gain and absorption regimes of the QD SOA. It confirms that in the QD SOA we examined, linear dispersion in the presence of the Kerr-like effect causes pulse compression, which counteracts the pulse peak suppression due to TPA, and also modifies the patterns which the coherent Rabi-oscillations imprint on the pulse envelope under both gain and absorption conditions. The inclusion of these effects leads to a better fit with experiments and to a better understanding of the interplay among the various mechanisms so as to be able to better analyze more complex future experiments of coherent light-matter interaction induced by short pulses propagating along an SOA.

I. INTRODUCTION

Active optical semiconductor waveguides, based on nano-metric gain media including quantum-dashes (QDashs) and quantum-dots (QDs) have been extensively studied over the last decade to have a better understanding of their dynamical properties, and to explore possible applications in fast and efficient communication and processing systems. QDs have also been the focus of fundamental research, where quantum-mechanical phenomena are routinely exploited in these so-called "artificial atoms", usually at cryogenic temperatures. Recently, the two research domains were bridged, as coherent light-matter interactions have been demonstrated in electrically driven, room-temperature semiconductor optical amplifiers (SOAs) based on QDashs operating at 1.55 μm wavelength[1] and later also in QD SOAs operating at 1.55 μm[2] and at 1.3 μm[3].

These coherent observations were made possible using ultra-fast techniques that can measure the complex envelope of an ultra-short light pulse after it had propagated through the waveguide of the SOA. The temporal width of the short pulses defined an interaction time shorter than the decoherence time of the medium, while the observation techniques provided the temporal resolution to observe details on time-scales shorter than the pulses themselves. The experimental findings were also supported by numerical calculations accounting for the interaction of an electromagnetic pulse with an active waveguide, approximated as a cascade of semi-classical homogeneous two-level systems[4] or a cascade of inhomogeneous ensembles of such two-level systems[2]. These comprehensive models were crucial for understanding how the coherent Rabi-oscillations, which the medium undergoes during its interaction with the light pulses, imprint a signature on the envelope of the propagating pulse, resulting in the temporal amplitude and phase profiles evident at the output of the device. These models enabled to identify qualitatively that coherent light-matter interactions were indeed responsible for the observed pulse-shapes. However, they did not account for any other propagation phenomena, i.e. dispersion, two-photon absorption (TPA) and the Kerr effect, and hence were somewhat inaccurate with respect to a few of the details of the observed signatures.

Non-resonant propagation effects were frequently observed in dynamical experiments when pulses propagated through semiconductor optical waveguides. Pump-Probe experiments with QDash SOAs at 1.5 μm[5, 6] and at 1 μm[7] as well as 1.5 μm[8] QD SOAs revealed that an intense pump pulse injects charge-carriers into the active region of the amplifier by TPA, and induces an increase in the gain which is experienced by a probe signal at the wavelength of the pump or any other wavelength within the gain spectrum. Since TPA absorbs the pulse peak more than its wings, it was also considered to be responsible for suppressing the pulse narrowing expected in self-induced-transparency (SIT)[1].

Zilkie et al. showed that the TPA process is accompanied by a Kerr-like effect and produces an instantaneously appearing negative line-enhancement factor in QD SOAs when biased to absorption[9]. A similar Kerr-like effect in a bulk semiconductor was reported by Siederdissen et al. to cause self-phase modulation, and to create "soliton-like" pulse-shapes, in combination with the linear dispersion[10]. Linear dispersion alone was measured, for example, in a QD SOA at 1.3 μm[11]. Romstad et al. have also demonstrated a complicated scenario for pulse propagation in bulk SOAs under different operating regimes involving non-resonant effects and nonadiabatic following[12], which called for further study to discriminate between their contributions. This was followed by a microscopic model for pulse-envelope propagation in bulk semiconductor[13] and a similar model was also reported for quantum-well SOAs[14]. However, none of these studies treated theoretically the non-resonant propagation mechanisms. Thus, the ability to discriminate these effects from the coherent Rabi-oscillations, by utilizing a wave-propagation numerical model, remained of great importance.

This paper reports such an investigation for QD SOAs operating at 1550 nm, which was conducted in two steps. First, we experimentally characterized, by cross frequency resolved optical gating (XFROG),[15] the pulse envelope at the output of a QD SOA when biased to its resonant transparency. In this regime, the coherent interactions were diminished, and any signature imprinted on the pulse was solely due to the non-resonant mechanisms, allowing their separate study. In the second step, we developed a new numerical model, which evaluates the effects of TPA, its accompanying Kerr-like effect, and linear dispersion in addition to the coherent light-matter interactions. This new model is based on a Finite-Difference Time-Domain (FDTD) algorithm avoiding any slowly-varying-envelope approximations, and therefore allows one to investigate extremely fast dynamics, shorter than an optical cycle. It enables to analyze the comprehensive response of the SOA alongside the pulse propagation in both the gain and absorption regimes of operation, allowing a separate investigation of the action of each of the mechanisms involved in shaping the pulse.

We found that in the transparency regime, the pulse experiences non-resonant absorption, as expected, and exhibits an instantaneous frequency profile which has the characteristics of a Kerr-like effect. For low intensities, the pulse showed a different, almost linear, chirp, which is dominated by linear dispersion. At sufficiently high energies, pulse narrowing was observed, indicating a soliton-like behavior. Using the newly developed numerical model, we also showed that the Kerr-like effect is crucial for the modeling of the experimentally observed signatures in both the gain and absorption regimes.

This paper is organized as follows. First, the experimental study of pulse propagation in a transparent QD SOA is presented in section 2. The numerical model is introduced in section 3 and section 4 presents the predictions of the model in the different operational regimes, and compares them with experimental results. Finally, section 5 is devoted to conclusions.

## II. PULSE PROPAGATION AT TRANSPARENCY

In order to discriminate experimentally between the signature of Rabi-oscillations and non-resonant propagation phenomena, we characterized the envelope of a 185 fs wide pulse after propagation through the QD SOA waveguide, when biased to transparency. In transparency, the probabilities for resonant absorption and for stimulated emission are almost equal, and hence the pulse does not induce any modification to the charge-carrier populations in the QDs, and no Rabi-floppings take place. Only non-resonant effects, such as dispersion, TPA, and its accompanying Kerr-like effect, can have a pronounced signature. Thus, it is possible to identify their imprints on the pulse envelope separately from the coherent phenomena.

The device we examined was a 1.5 mm long edge emitting SOA comprising four layers of self-assembled InAs QDs placed between InGaAlAs barriers, grown on an InP substrate[16]. The bias dependent amplified spontaneous emission spectra are shown in Fig. 1, exhibiting a 70 nm wide inhomogeneously broadened gain spectrum. The excitation pulses were filtered from the output of a Toptica FemtoFiber Pro fiber laser, with maximum pulse energy of about 250 pJ (coupled to the SOA). The spectral shape of the excitation pulse is also shown in Fig. 1, exhibiting a bandwidth of about 20 nm (FWHM) centered at 1540 nm. Thus, and because of the inhomogeneous gain broadening, the true transparency was only approximately achieved, as some parts of the pulse spectrum always experience some absorption or amplification. Nonetheless, transparency was defined at the bias level where the effects of stimulated emission and absorption are minimal, by a simple single-wavelength pulsed pump-probe experiment. A bias of approximately 95 mA was found to be the transparency point since at this level the transmission of the probe pulse was not affected by the presence of the pump pulse, preceding it by a few hundreds of femtoseconds.

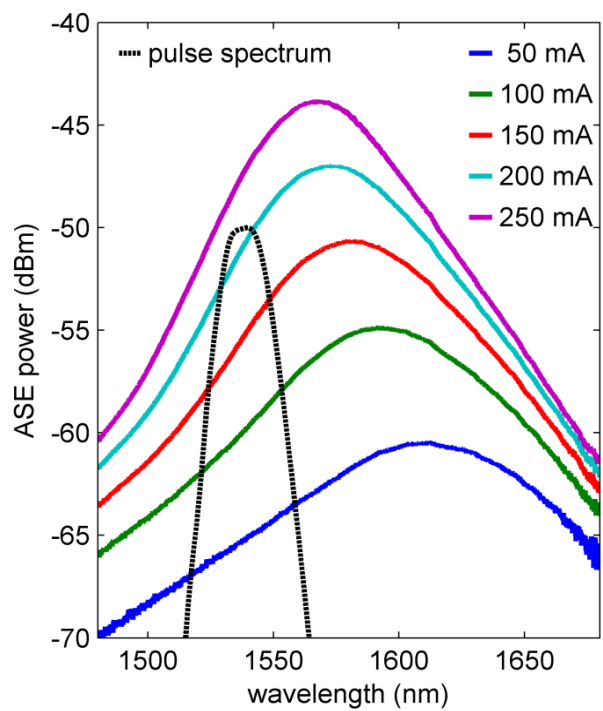

FIG. 1. Bias dependent emission spectra of the SOA, together with the spectrum of the excitation pulse used in the experiments.

Next, we used the XFROG system to analyze the modifications induced on the pulse complex envelope in the above transparency point, for various input pulse energies. The results are summarized in Fig. 2 (a) and (b), showing the temporal intensity and instantaneous frequency (chirp) profiles, respectively.

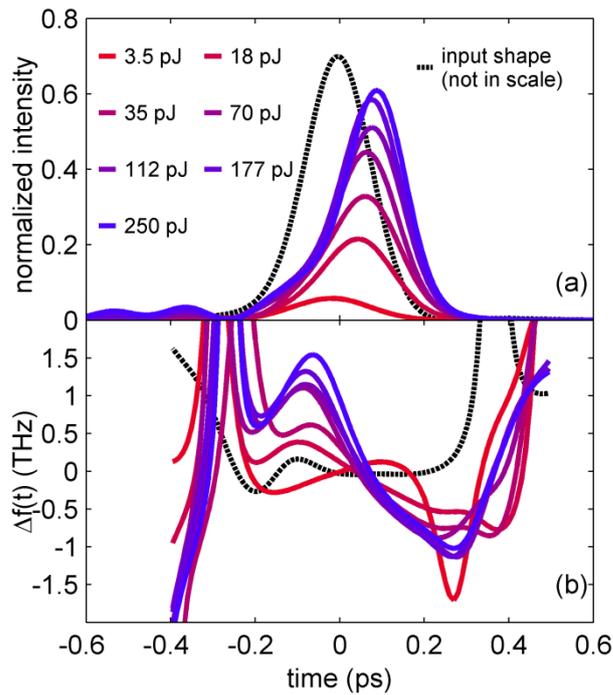

FIG. 2. Transparency regime. (a) Measured time-dependent intensity profiles of the output pulses for various input energies. (b) Instantaneous frequency profile of the output pulses. The dashed curve presents the input pulse profiles. Its intensity is not plotted in the same scale as the output pulses.

The measured intensity profiles show that as the input energy increases, the output pulses peaks become asymmetric and delayed. The peak intensities fail to follow the increase in the input energy, as also clearly presented in Fig. 3 (a) (blue dots). Since the pulses are not broadened at the same time, this trend of the peak intensity indicates that the pulses experience an intensity dependent absorption upon propagation along the waveguide.

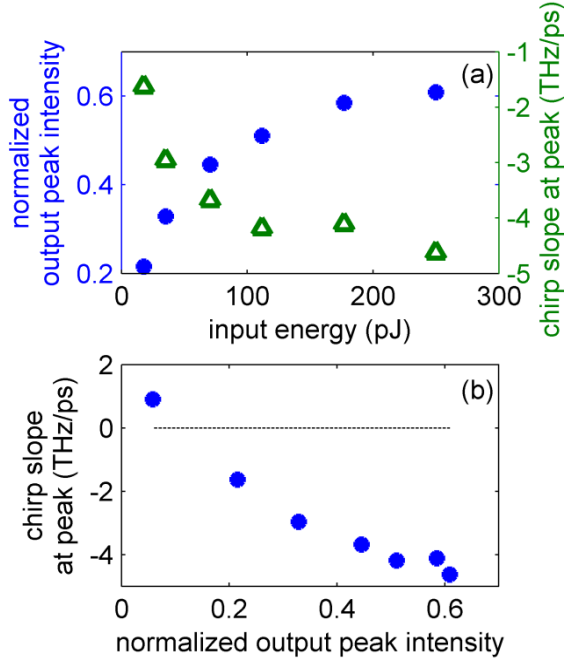

FIG. 3. Transparency regime. (a) Normalized peak intensity at the output of the waveguide (blue dots), and chirp slope (green triangles) as function of the pulse energy at the input. (b) chirp slope as function of the normalized output peak intensity.

The measured chirp profiles reveal a transition between two regimes. For a powerful, 250 pJ, input pulse, where the output intensity profile is the most augmented one in Fig. 2 (a), the chirp profile shows a steep instantaneous frequency decrease. As the input energy is lowered, this chirp slope decreases (in its absolute value), down to a point where for the weak, 3.5 pJ, pulse, it exhibits a positive slope. The trend of the negative chirp slope with the input energy resembled that of the output peak intensity with the input energy as shown in Fig. 3(a) (green triangles). However, when plotted against the output peak intensity, the chirp slope changes almost linearly, as seen in Fig. 3 (b). This trend implies an instantaneous, intensity dependent, phase modulation, namely a Kerr-like effect. The presence of such a phenomenon has been reported in several past cases[9, 10] in the context of TPA. Hence, we conclude that the observed effects are due to TPA (responsible for the intensity dependent absorption) with an accompanying Kerr-like effect.

As the pulse intensity is reduced, these effects decay, and modifications to the pulse are dominated by linear dispersion, which creates the positive chirp slope. Opposite signs of the chirp, caused by dispersion and the Kerr-like effect, imply that the pulse narrowing at high energies is in fact a manifestation of "soliton-like" propagation, similar to the one reported in Ref. 10.

Thus, the experiments in the transparency regime highlight the significance of TPA, its accompanying Kerr-like effect, and to a lesser extent that of the linear dispersion, in shaping the pulses which propagate in the SOA. Deciphering the combined effect of the various propagation effects and the coherent Rabi-oscillations evident in higher or lower bias levels, requires a proper model. Accordingly, such a new comprehensive FDTD model that considers all these phenomena, and allows a thorough study of their collective signature, is introduced in the next section.

### III. NUMERICAL MODEL

The proposed model solves Maxwell's equations governing the electromagnetic pulse propagation, where the response of the medium is expressed through the induced polarization term. Each phenomenon is therefore considered by its particular contribution to the induced polarization. We invoke a full-wave FDTD numerical algorithm that allows examination of the co-evolution of the electromagnetic wave and any other fast dynamical mechanism, such as the evolution of a two-level system, without using any rotating-wave or slowly-varying-envelope approximations.

The present model extends our previous models of coherent light-matter interactions[2,4]. The electromagnetic field is assumed to propagate as a TEM mode along the Z-axis of the amplifier, which is modeled in turn as a cascade of ensembles of quantum-mechanical two-level systems, expressing the inhomogeneous spectral broadening of the self-assembled QD medium. The dynamics of these two-level systems are treated by solving the Schrödinger equation in the density-matrix formalism. Each two-level system describes a ground state in a QD which is fed, incoherently, with charge carriers from an excited state belonging to the same QD. Charge carriers are captured in, or escape out of these QDs to a carrier reservoir residing at higher energy levels, which is fed by the external current supply. All these incoherent dynamics are evaluated using a set of rate equations. Technically speaking, Maxwell's and Schrödinger's equations are treated with a central difference discretization, and the slower rate equations are propagated with a forward difference discretization.

The calculation of pulse propagation is established, as already mentioned, on the solution of Maxwell's curl equations, for an electric field polarized along the X-axis, and a magnetic field pointing in the Y direction:

$$\begin{cases} \dfrac{\partial E_x}{\partial z} = -\mu_0 \dfrac{\partial H_y}{\partial t} \\ -\dfrac{\partial H_y}{\partial z} = \dfrac{\partial D_x}{\partial t} \end{cases} \quad (1)$$

Where $E_x$, $D_x$, and $H_y$ are the electric field, electric displacement, and magnetic field components, respectively, $\mu_0$ is the permeability of the vacuum. The displacement is related to the electric field and to the induced polarization $P_x$:

$$D_x = \varepsilon_0 \varepsilon_\infty E_x + P_x \quad (2)$$

with $\varepsilon_0$ being the permittivity of the vacuum, and $\varepsilon_\infty$ is the dielectric constant for an infinite frequency. In every time-step of the FDTD algorithm, Eq. 1 is evaluated to provide the values for $H_y$ and $D_x$ along the amplifier. These are used to calculate the various polarization elements, which in turn enable to calculate $E_x$ based on Eq. (2), and proceed to the next time step.

The polarization $P_x$ is comprised of several contributions:

$$P_x = P_{dispersion} + P_{Kerr} + P_{TPA} + P_{QD} + P_{plasma} \quad (3)$$

$P_{dispersion}$ accounts for the linear (with the field) contribution. $P_{TPA}$ and $P_{Kerr}$ represent the contributions of the TPA of the wave and of the accompanying Kerr-like effect, respectively. The interaction with the QDs and their charge carriers is folded in $P_{QD}$ and $P_{plasma}$, determining the radiation of the two-level systems, and the changes in refractive index due to the carrier population at the various energy levels.

The linear dispersion is introduced phenomenologically, by driving a lossless Lorentz oscillator,[17,18] avoiding frequency dependent absorption arising from this contribution:

$$\dfrac{\partial^2 P_{dispersion}}{\partial t^2} + \omega_L^2 P_{dispersion} = \varepsilon_0 \chi_L \omega_L^2 E_x \quad (4)$$

Here, $\omega_L$ is the resonant frequency of the Lorentz oscillator, and $\chi_L$ is the oscillator strength. Both serve to determine the desired refractive index and the desired degree of group velocity dispersion (GVD). This is performed by transforming (4) to the frequency domain, obtaining:

$$P_{dispersion}(\omega) = \frac{\varepsilon_0 \chi_L \omega_L^2 E_x}{\left(\omega_L^2 - \omega^2\right)} \tag{5}$$

Hence, the nominal dielectric constant at the central excitation frequency $\omega$ is given by (including the response of the medium at infinite frequency $\varepsilon_\infty$):

$$\varepsilon(\omega) = \varepsilon_\infty + \frac{\chi_L \omega_L^2}{\omega_L^2 - \omega^2} \tag{6}$$

The GVD is then expressed by the second derivative, with respect to $\omega$, of the propagation coefficient $\beta(\omega) = \frac{\omega}{c}\sqrt{\varepsilon(\omega)}$ (c is the speed of light in vacuum). Choosing a nominal dielectric constant of 12.25, representing this property of semiconductors at optical frequencies[19], and fixing the GVD to a desired positive value, one is able to solve for the parameters $(\chi_L, \omega_L)$ of the Lorentz oscillator. Then, the dynamics of this oscillator are evaluated in each time-step of the calculation process, using the known values of the field variables $P_{dispersion}$ and $D_x$, according to the following discretization scheme[17]:

$$\frac{P_{dispersion}^{n+1} - 2P_{dispersion}^n + P_{dispersion}^{n-1}}{\Delta t^2} + \frac{\omega_L^2 \left(P_{dispersion}^{n+1} + P_{dispersion}^{n-1}\right)}{2} = \frac{\omega_L^2 \chi_L \left(D^{n+1} + D^{n-1} - P_{dispersion}^{n+1} - P_{dispersion}^{n-1}\right)}{2\varepsilon_\infty} \tag{7}$$

$\Delta t$ is the time increment used in the simulation, and the superscripts n-1, n, n+1 denote the previous time-step, the current time, and the next time-step, respectively. If negative GVD values are needed, a Drude model can similarly be applied instead of the Lorentz oscillator.

The treatment of the non-linear terms starts by formulating the relation between the TPA susceptibility and the strength of the Kerr effect, in an α-factor relation, which is described in the frequency domain by:

$$P_{Kerr+TPA}(\omega) = \varepsilon_0 (1 - j\alpha_{TPA}) \chi_{TPA} E_x(\omega) \tag{8}$$

Where $j = \sqrt{-1}$, and $\alpha_{TPA}$ is the factor relating the effect of TPA on the refractive index, to its effect on the amplitude of the wave[9]. $\chi_{TPA}$ is the TPA susceptibility. In the frequency domain, it is related to the TPA efficiency $\beta_{TPA}$ (deifned by Beer's law for the light intensity: $\frac{\partial I}{\partial z} = -\beta_{TPA} I^2$) by[18]:

$$\chi_{TPA} = -\frac{c^2 \varepsilon_0 n_0^2 \beta_{TPA}}{2j\omega} |E_x|^2 \tag{9}$$

with $n_0$ being the nominal refractive index. Since one phenomenon has an imaginary susceptibility and the other has a real one, different treatments are required to fit them into the FDTD simulator that deals with real polarizations only. The Kerr effect simply modifies the refractive index of the medium, and its induced polarization is real, formulated as[18]:

$$P_{Kerr} = \varepsilon_0 \frac{\alpha_{TPA} \varepsilon_0 c^2 n_0^2 \beta_{TPA} |E_x|^2}{2\omega} E_x \tag{10}$$

This is calculated at the central frequency of the wave. In principle, more complicated effects such as self-steepening may be considered. In that case, the Kerr polarization must include an independent dynamics driven by the wave[18], in a similar fashion to what has been used for the case of linear dispersion. These complications have been avoided in the present model.

The contribution of the TPA term must be considered after performing a time derivative of (2)[18]:

$$\frac{\partial D_x}{\partial t} = \varepsilon_0 \varepsilon_\infty \frac{\partial E_x}{\partial t} + \frac{\partial P_{dispersion}}{\partial t} + \frac{\partial P_{Kerr}}{\partial t} + \frac{\partial P_{QD}}{\partial t} + \frac{\partial P_{plasma}}{\partial t} + P_{TPA} \tag{11}$$

Where:

$$P_{TPA} = \frac{c^2 \varepsilon_0^2 n_0^2 \beta_{TPA} |E_x|^2}{2} E_x \qquad (12)$$

The polarization contributed by the two-level systems is calculated from the coherence terms in their density matrices[4]. The dynamic behavior is obtained from the dynamics of Schrödinger's equation. Although it affects the wave amplitude, it is a real polarization term, and therefore it too was treated with a time derivation in (11).

Finally, the contribution of the plasma effect is dealt with by introducing the proper polarization contribution, considering the modifications made to the dielectric constant in the medium:

$$P_{plasma} = \varepsilon_0 \Delta\varepsilon_{plasma} E \qquad (13)$$

This change is defined phenomenologically by the population of charge carriers in all the energy levels:

$$\Delta\varepsilon = C_{11} \cdot 2 \sum_{ensemble} N_d^i \rho_{11}^i \\ + C_{22} \cdot 2 \sum_{ensemble} N_d^i \rho_{22}^i \\ + C_{ex} \sum_{ensemble} N_{ex}^i \\ + C_{res} N_{res} + C_{h\_res} h_{res} \qquad (14)$$

Where $C_{res}$, $C_{ex}$, $C_{11}$, $C_{22}$, and $C_{h\_res}$ are phenomenological coefficients describing the change in the dielectric constant due to carrier populations in the electron reservoir, excited states of the QDs, upper states (of the two-level systems), lower states and the hole reservoir, respectively. In (14) these are multiplied respectively by the populations $N_{res}$, $\sum_{ensemble} N_{ex}^i$, $2\sum_{ensemble} N_d^i \rho_{11}^i$, $2\sum_{ensemble} N_d^i \rho_{22}^i$, and $h_{res}$, where summations are performed in order to account for the entire QD inhomogeneously broadened spectrum. $N_d^i$ is the QD volume density for the i-th sub-group of QDs, and $\rho_{11}^i$ ($\rho_{22}^i$) is the population probability of the upper (lower) state in the corresponding two-level system. Adding this polarization to (11) yields:

$$\frac{\partial D_x}{\partial t} = \varepsilon_0 \varepsilon_\infty \frac{\partial E_x}{\partial t} + \frac{\partial P_{dispersion}}{\partial t} + \frac{\partial P_{Kerr}}{\partial t} + \frac{\partial P_{QD}}{\partial t} + \varepsilon_0 \frac{\partial \left(\Delta\varepsilon_{plasma} E_x\right)}{\partial t} + P_{TPA} \qquad (15)$$

This relation is evaluated in each time step in order to calculate the electric field for the next step. The difference equation obtained by discretizing (15) is advanced iteratively[18].

Unlike the wave propagation procedure, which was significantly altered from the previous models,[2, 4] the charge-carrier dynamics are only slightly modified, so as to accommodate the additional charge carriers injected by TPA. It has been assumed that they are injected into a very high energy level with an infinite density of states, so that population inversion cannot occur there. From this level, the carriers relax to the electron or hole reservoirs, or recombine, obeying the following rate equations:

$$\begin{cases} \dfrac{\partial N_{TPA}}{\partial t} = I_{TPA} - \left(1 - \dfrac{N_{res}}{D_{res}}\right)\dfrac{N_{TPA}}{\tau_{TPA\_relax}} - \dfrac{N_{TPA}}{\tau_{TPA\_rec}} \\ \dfrac{\partial h_{TPA}}{\partial t} = I_{TPA} - \left(1 - \dfrac{h_{res}}{D_{res}}\right)\dfrac{1}{\tau_{h\_TPA\_relax}}h_{TPA} - \dfrac{N_{TPA}}{\tau_{TPA\_rec}} \end{cases} \quad (16)$$

$N_{TPA}$ and $h_{TPA}$ are the electron and hole population in the TPA injection levels. $D_{res}$ is the density of states in the carrier reservoirs which the injected carriers relax into, with the time constants $\tau_{TPA\_relax}$ and $\tau_{h\_TPA\_relax}$, for the injected electrons and holes, respectively. $\tau_{TPA\_rec}$ is the time-scale of direct recombination of the charge carriers in those levels. $I_{TPA}$ is the generation rate of carriers by TPA. It is related to the light intensity I and to the electric field by[18]:

$$I_{TPA} = \dfrac{\beta_{TPA}}{2\hbar\omega}I^2 = \dfrac{\beta_{TPA}c^2\varepsilon_0^2 n^2}{8\hbar\omega}|E_x|^4 \quad (17)$$

This concludes the introduction of our comprehensive numerical model.

IV. SIMULATION RESULTS AND DISCUSSION

The model described in section III was used to examine the imprints of the combined effects of coherent light-matter interactions and non-resonant propagation phenomena on the envelope of a 185 fs wide, transform-limited, pulse, launched at the input of the calculation space. Since many phenomenological parameters were involved, and due to the large computational complexities of the calculation process, the simulations served only to follow qualitatively the trends observed in the experiments. To that end, we modeled a shorter propagation length, with relatively strong parameters for the various mechanisms. The parameters which had to be tuned for the task were the bias level of the device, the pulse input energy, the GVD, TPA efficiency $\beta_{TPA}$, the Kerr-like parameter $\alpha_{TPA}$, the dipole moments for the two-level-systems, the overall density of QDs, and the various α-parameters of the charge-carrier populations. All these have a direct impact on the time-scale of the pulse itself, and affect its reshaping. The dipole-moment was tuned to obtain pulse areas of roughly 4π for the maximum considered pulse energy (800 pJ before coupling losses). The QDs' density was fixed to a value that enabled to observe their signature on the pulse shape, accumulated along the short waveguide we modeled. Other parameters which had a lesser effect on the pulse-shape were kept constant throughout the investigation. Their values, following our previous experiences with similar models,[1] are listed in Table I.

Reconstructing the different signatures on the pulse envelope required to identify the simulation bias levels which corresponded to the bias points in the experiments, and then tuning the parameters governing the propagation and the carrier effects. This tuning followed a few guidelines in order to keep the model physically reasonable: TPA efficiency and its accompanying Kerr effect were fixed for all the bias points,[9] the GVD value was allowed to change monotonically with the bias level[11, 20], and the effect of carrier population on the refractive index of the medium was assumed to be stronger in the gain regime compared to the absorption regime[9].

The results of the calculations are described in three sub-sections, concentrating on operating at the transparency point, in the gain regime and finally in absorption. In each sub-section, the results are compared to experimental observations, and the origins of the various signatures are discussed.

| Parameter | Value |
|---|---|
| 1. Active region geometry | |
|    a. Length (μm) | 300 |
|    b. Width (μm) | 4 |
|    c. Number of QD layers | 5 |
|    d. Layer thickness (nm) | 10 |
|    e. Confinement factor (%) | 0.25 |
|    f. Nominal refractive index | 3.5 |
| 2. Electromagnetic stimulus | |
|    a. Central wavelength (nm) | 1540 |
|    b. FWHM (fs) | 185 |
| 3. Resonant gain medium | |
|    a. Dipole moment (C·m) | $0.8 \cdot 10^{-28}$ |
|    b. Peak wavelength (nm) | 1580 |
|    c. Inhomogeneous broadening (nm) | 70 |
|    d. Homogeneous broadening (nm) | 25 |
|    e. Number of sub-levels | 44 |
| 4. Rate equations parameters | |
|    a. Total QD density ($m^{-3}$) | $4 \cdot 10^{23}$ |
|    b. Carrier reservoirs' density of states ($m^{-3}$) | $2 \cdot 10^{25}$ |
|    c. Carrier reservoir wavelength (nm) | 1450 |
|    d. Ground-Exited states energy separation (nm) | 50 |
|    e. Recombination times (ns) | 0.4 |
|    f. Electron capture time into the QDs (ps) | 1 |
|    g. Excited-Ground state relaxation time (fs) | 100 |
|    h. Hole capture time into the QDs (fs) | 100 |
|    i. Hole escape time from the QDs (fs) | 100 |
|    j. Relaxation time of the TPA electrons (ps) | 3 |
|    k. Relaxation time of the TPA holes (ps) | 3 |

Table I. List of the constant parameters used in the simulations.

### A. Transparency

Transparency point was identified in the model as the bias level in which the two-level systems in resonance with the pulse central wavelength had equal occupation probabilities in both their eigenstates, in absence of any electromagnetic excitation. For the active region geometry, gain spectrum properties, and QD density listed in Table I, this was achieved at 42 mA. To provide a positive chirp slope for weak pulses, similar to what was observed for 3.5 pJ pulses (see Fig. 2), a positive GVD was assumed in the simulation. For more intense pulses, TPA truncated the pulses' peaks, effectively broadening them in time. However, the accompanying Kerr effect, which introduced a negative $\alpha_{TPA}$ value, created a positive chirp on the leading edge of the pulses, and a negative chirp on their trailing edges. Together with the dispersion, it acted to compress the pulses into a "soliton-like" structure. A proper balance between these effects was required for the reconstruction of the observed pulse narrowing. Furthermore, the coefficients of the carriers' plasma effect were introduced, keeping their values in the carrier reservoirs to be higher than those of the QD ground states. As a consequence, the carriers which were generated by TPA also contributed a chirp via the plasma effect. For positive α-parameters, this effect resulted in a slight increase of the instantaneous frequency of the pulse along its trailing slope, effectively assisting the linear dispersion. The results, obtained with the parameters listed in Table II, for input pulse energies between 10 pJ to 800 pJ, are plotted in Fig. 4 (a) and (b), showing the time-dependent intensity and chirp profiles, respectively, of the output pulses. The inset of Fig. 4 (a) presents the pulse full widths at half maximum (FWHM) as function of their input energies (Note the logarithmic scale of the energy axis). It shows that the weak pulses are broadened by dispersion, while the powerful pulses are compressed, as demonstrated experimentally. The intensity profiles in Fig. 4 (a) demonstrate the effect of TPA in diminishing the pulse peak with respect to its initial energy. Figure 4 (b) presents the evolution of the instantaneous frequency traces from the rising, dispersion dominated, profiles of the weak pulses, to the decreasing, Kerr effect dominated, profiles for the intense pulses. Within the present computational limitations, these results resemble the experimental findings well.

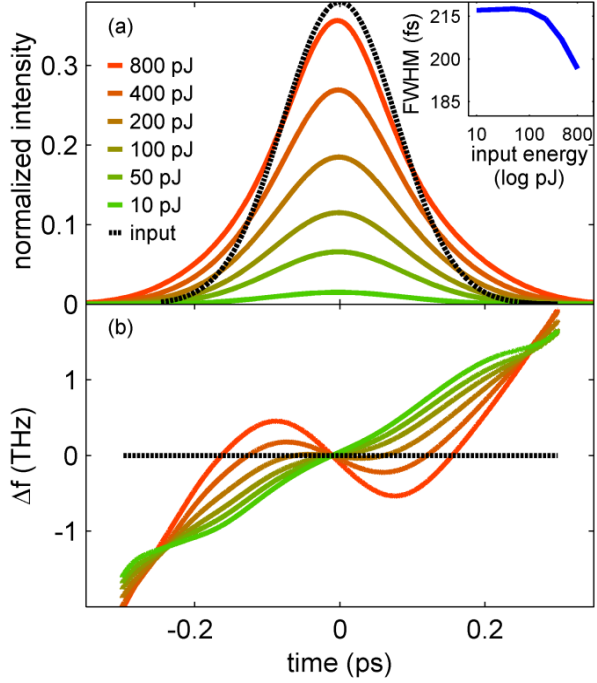

FIG. 4. Pulse propagation at transparency. (a) Calculated output pulse time-dependent intensity for various input energies. Inset: corresponding pulse widths versus input energy. (b) Corresponding calculated instantaneous frequency profiles.

| Parameter | Value |
| --- | --- |
| 1. propagation effects | |
| a. $\beta_{TPA}$ (GW/cm$^2$) | 1000 |
| b. $\alpha_{TPA}$ | -3 |
| c. GVD (fs$^2$/mm) | 20000 |
| 2. plasma effect coefficients | |
| a. $C_{res}$ (m$^3$) | $0.25 \cdot 10^{-26}$ |
| b. $C_{ex}$ (m$^3$) | $0.15 \cdot 10^{-26}$ |
| c. $C_{11}$ (m$^3$) | $0.05 \cdot 10^{-26}$ |
| d. $C_{22}$ (m$^3$) | $0.08 \cdot 10^{-26}$ |
| e. $C_{h\_res}$ (m$^3$) | $0.3 \cdot 10^{-26}$ |

Table II. Simulation parameters for balancing GVD, TPA, Kerr effect and the plasma effect at the transparency bias of 42 mA.

## B. Gain regime

In the next stage of the investigation, the effects of GVD, TPA and its accompanying Kerr effect were studied in the gain regime (using a simulated bias of 100 mA), where coherent light-matter interactions are pronounced. At this bias level, the same pulses which were tested at transparency, triggered up to two cycles of Rabi-oscillations in the resonant two-level systems. In order to clarify the role of these Rabi-oscillations in shaping the pulse envelopes, as compared to the non-resonant effects, we considered two cases, one with and the second without the TPA and the accompanying Kerr effect contributions. The predicted output pulse-shapes for various input energies, obtained using the parameters summarized in Table III, and compared with an experimental measurement of similar pulses which had propagated through the SOA under gain conditions, are plotted in Fig. 5. It presents the time-dependent intensity profiles of the predicted pulses without accounting for the TPA and Kerr effects (Fig. 5 (a)), of the predicted pulses with these effects included (Fig. 5 (c)), and of the experimental measurement (Fig. 5(e)). Figure 5 (b), (d), and (f) show the corresponding instantaneous frequency profiles.

| Parameter | Value |
|---|---|
| 1. propagation effects | |
| a. $\beta_{TPA}$ (GW/cm$^2$) | 1000 |
| b. $\alpha_{TPA}$ | -3 |
| c. GVD (fs$^2$/mm) | 2000 |
| 2. plasma effect coefficients | |
| a. $C_{res}$ (m$^3$) | $0.3 \cdot 10^{-26}$ |
| b. $C_{ex}$ (m$^3$) | $0.2 \cdot 10^{-26}$ |
| c. $C_{11}$ (m$^3$) | $0.05 \cdot 10^{-26}$ |
| d. $C_{22}$ (m$^3$) | $0.1 \cdot 10^{-26}$ |
| e. $C_{h\_res}$ (m$^3$) | $0.4 \cdot 10^{-26}$ |

Table III. Simulation parameters for balancing GVD, TPA, Kerr effect and the plasma effect at 100 mA bias, availing gain regime.

In Fig. 5 (a), the intensity profiles reveal the evolution of the oscillations imprinted by the coherent interaction with the two-level systems. The appearance of a second lobe for the 50 pJ pulse testifies that the pulse has triggered a complete Rabi-cycle, and a second one is starting. When the pulse intensity increases further, this second cycle advances, and the imprinted oscillations in the amplitude become clear. Moreover, as the pulse intensity rises, the first peak advances gradually to earlier times, consistent with the rise in the Rabi-frequency. A matching pattern is visible also in the chirp profiles of Fig. 5 (b). Each Rabi-cycle is accompanied by a temporary red-shift, caused by the plasma effect of the charge carriers on the refractive index of the medium[1]. However, this pattern is significant only along the first Rabi-cycle, which is explained as follows. The fast carrier replenishment into the QDs from the carrier reservoir causes depletion of the reservoir. Since the plasma effect of the reservoir is strong, it creates a pronounced red-shift on the leading edge of the pulse. On the other hand, this relaxation quickly averages out the population oscillations in the carrier reservoir, and so the second Rabi-cycle is manifested by a minor red-shift in the chirp profile, as the population of the QDs affect the chirp much less than that the reservoir.

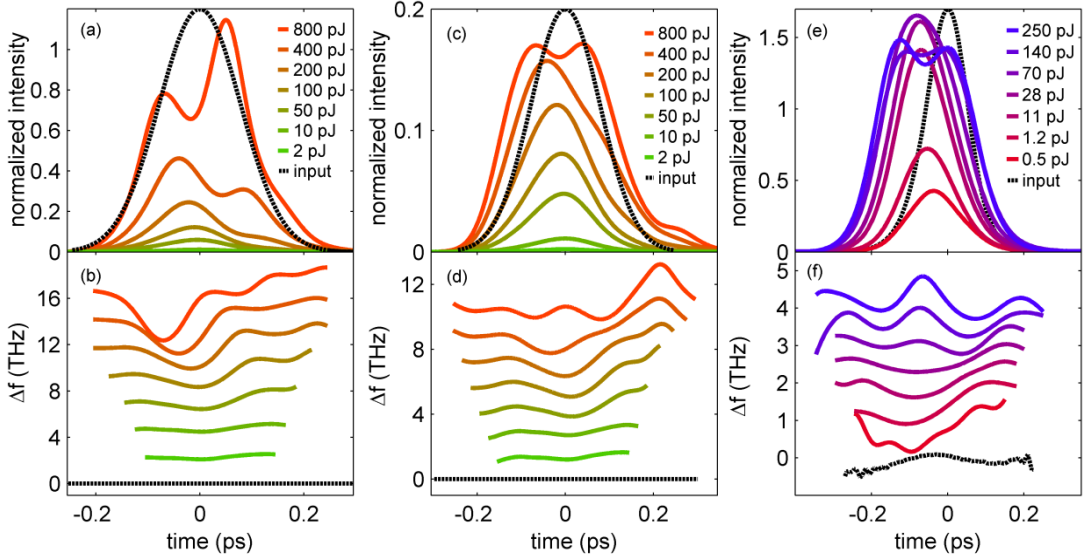

FIG. 5. Pulse propagation in gain regime. (a) Calculated time dependent intensity of the output pulses, not accounting for TPA and Kerr effect. (b) The corresponding instantaneous frequency patterns. (c) Calculated time dependent intensity of the output pulses, including TPA and Kerr effect. (d) The corresponding instantaneous frequency patterns. (e) Measured time dependent intensity of the output pulses from the real SOA. (f) The corresponding instantaneous frequency patterns.

Introducing the TPA and its accompanying Kerr effect, while keeping the other parameters fixed, resulted in the evolution presented in Fig. 5 (c) and (d). Similar to the case of transparency, the TPA itself suppresses and clamps the peak intensities of the high energy pulses. Thus, the pulse areas are effectively reduced, and their coherent interactions with the two-level systems are weakened, preventing the appearance of the second lobe at the lower energies. Nevertheless, Fig. 5 (c) still shows clearly the intensity oscillations due to the coherent interaction, when the pulse intensity is sufficiently high. The Kerr effect acts to raise the instantaneous frequency whenever the pulse intensity rises, and lower it when the pulse intensity reduces. Together with the linear dispersion this effect balances the chirp induced by the charge carriers, and also slightly compresses the pulse lobes, bringing them closer together. Thus, the overall evolution of the pulse envelope is milder compared with the case without TPA. For the weaker pulses, the lower Rabi-frequency, together with the "soliton-like" pulse compression create narrower envelopes than for the case with TPA ignored. When the pulse area is sufficiently high, it splits into two lobes, which drift apart considerably (but not as much as without TPA), and produce two corresponding red-shifts which are of similar depths, as demonstrated in Fig. 5 (d). The pulse envelopes, measured at the output of the QD SOA under a bias of 250 mA, are shown in Fig. 5 (e) and (f). The figures exhibit a qualitatively similar evolution. The oscillations in the intensity profile (Fig. 5 (e)) do not grow gradually but rather appear as the input pulse energy crosses a certain value. The peak intensity is clamped, especially when the input intensity produces the coherent oscillations. The corresponding chirp profile, in Fig. 5 (f), also presents mild features, with a single red-shift feature turning, upon crossing the coherent break-up "threshold", into two similar valued red-shifts, which are torn apart from each other. We therefore conclude that in the gain regime, the new model that combines the coherent Rabi oscillations and the non-resonant effects improves the predictions of the pulse shapes at the SOA output, and enables to discriminate among the contributions from each of the mechanisms.

### C. Absorption regime

Finally, absorption conditions were modeled by applying a 10 mA bias, and the non-resonant parameters listed in Table IV. Here again, the results of the full model were compared to a calculation ignoring the TPA and the Kerr effect, and to an experimental measurement. The comparisons are depicted in Fig. 6, which present the time-dependent intensity profiles of the predicted pulses without accounting for the

TPA and Kerr effects (Fig. 6 (a)), of the predicted pulses with these effects included (Fig. 6 (c)), and of the experimental measurement Fig. 6 (e)). Figure 6 (b), (d), and (f) show the corresponding instantaneous frequency profiles.

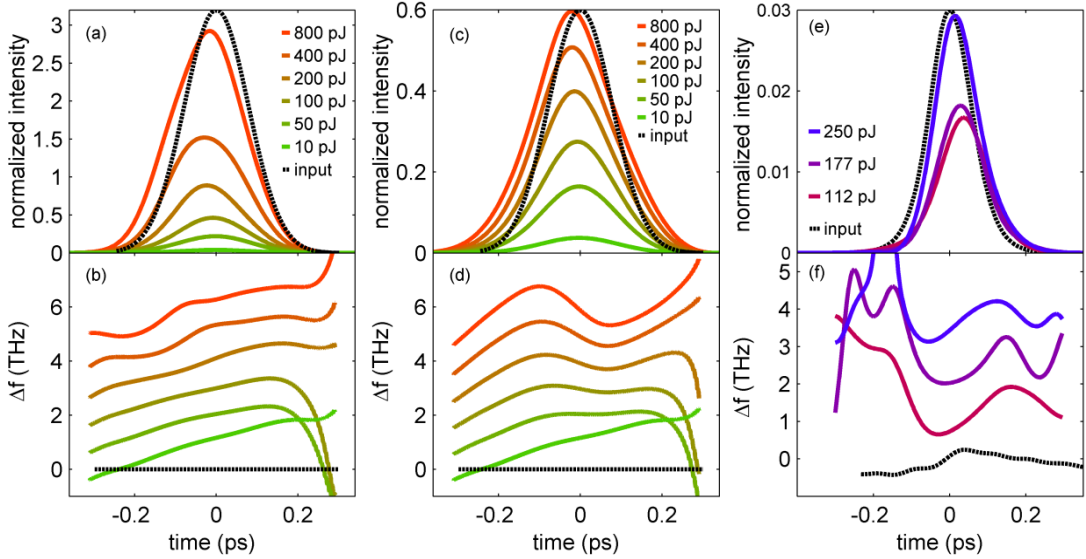

FIG. 6. Pulse propagation in absorption regime. (a) Calculated time dependent intensity of the output pulses, not accounting for TPA and Kerr effect. (b) The corresponding instantaneous frequency patterns. (c) Calculated time dependent intensity of the output pulses, including TPA and Kerr effect. (d) The corresponding instantaneous frequency patterns. (e) Measured time dependent intensity of the output pulses from the real SOA. (f) The corresponding instantaneous frequency patterns.

| Parameter | Value |
|---|---|
| 1. propagation effects | |
| a. $\beta_{TPA}$ (GW/cm$^2$) | 1000 |
| b. $\alpha_{TPA}$ | -3 |
| c. GVD (fs$^2$/mm) | 20000 |
| 2. plasma effect coefficients | |
| a. $C_{res}$ (m$^3$) | $0.15 \cdot 10^{-26}$ |
| b. $C_{ex}$ (m$^3$) | $0.1 \cdot 10^{-26}$ |
| c. $C_{11}$ (m$^3$) | $0.02 \cdot 10^{-26}$ |
| d. $C_{22}$ (m$^3$) | $0.05 \cdot 10^{-26}$ |
| e. $C_{h\_res}$ (m$^3$) | $0.2 \cdot 10^{-26}$ |

Table IV. Simulation parameters for balancing GVD, TPA, Kerr effect and the plasma effect at 10 mA bias, availing absorption regime.

Since the pulses are absorbed, their areas are in general diminished so that the common shape of their intensity profiles has a single peak, without any oscillations. At best, some pulse compression is predicted, as seen in Fig. 6 (a) and (c), and confirmed in the measured traces (Fig. 6 (e)). The only visible difference between the calculated pulse-shapes with TPA and without it is the peak suppression due to the intensity dependent absorption.

The instantaneous frequency profiles, however, provide a more powerful evidence for the role of the non-resonant effects. The calculations which disregard TPA predict that the absorption of the pulse, generating charge carriers in the SOA, induces a positive chirp along the leading edge of the pulse, seen in Fig. 6 (b). Even for intense pulses, which trigger coherent self-induced transparency (SIT), the chirp is mainly positive, again due to the fast escape of charge carriers from the two-level systems. The introduction of the TPA with it accompanying Kerr effect creates a mechanism that induces a negative effect on the chirp profile[9]. It induces the temporal decrease in the instantaneous frequency shown in Fig. 6 (d). Since this chirp is opposite to that induced by the linear dispersion of the medium it also leads to some pulse compression.

This red-shift is clearly evident in the experimental findings in Fig. 6 (f), and therefore further confirms the need to include TPA and the Kerr effect in the model. The observed compression of the intensity profile for 250 pJ input energy (Fig. 6 (e)) is explained by the SIT mechanism assisted by the focusing action of the Kerr-like effect and the dispersion.

## V. CONCLUSION

In conclusion, we studied the impact of non-resonant propagation effects on the envelope of an ultra-short pulse as it propagates through a QD SOA and experiences coherent light-matter interactions such as Rabi-oscillations.

XFROG characterization of the pulse complex envelope at the output of the SOA under transparency conditions (where the coherent interactions are diminished) revealed that linear dispersion, TPA, and a Kerr-like effect contribute significantly in shaping the pulse envelope. In order to discriminate these effects in the gain and absorption regimes as well, a new comprehensive numerical model describing the pulse propagation was developed. This model integrates our previously developed FDTD numerical simulation tools describing the propagation of a pulse through a chain of quantum mechanical two-level systems with additional FDTD schemes that model the non-resonant propagation effects of TPA, Kerr, and GVD. The new model reconstructed qualitatively the pulse shapes and their evolution with the pulse input intensity at transparency, gain, and absorption. While the coherent Rabi-oscillations were undoubtedly found responsible for the observed pulse break-up at high input energies in the gain regime, the focusing action of the linear dispersion combined with the Kerr effect was essential for producing key features such as: (i) At transparency, "soliton-like" propagation and pulse narrowing at high intensities as observed in the experiments. (ii) In gain regime, constraining the pulse break-up and clamping of the peak intensity, as well as the reproduction of two equally deep red-shifts which cannot be reconstructed otherwise. (iii) In the absorption regime, the red-shifts observed in the experiments were predicted only by the action of the Kerr effect.

Conclusively, this investigation provides a better understanding of the mechanisms shaping the pulse while propagating in the SOA, and will serve to predict and delineate more accurately the result of future experiments.

ACKNOWLEDGMENTS

This work was partially supported by the Israeli Science Foundation. O. Karni would like to thank the financial support provided during this research by the Daniel Fellowship, and by the Russel Berrie Fellowship.